# A Taxonomy of Performance Prediction Systems in the Parallel and Distributed Computing Grids

SENA SENEVIRATNE & DAVID C. LEVY
*Computer Engineering Lab.*
*School of Electrical and Information Engineering*
*The University of Sydney, Australia*

RAJKUMAR BUYYA
*Cloud Computing and Distributed Systems Lab.*
*Department of Computing and Information Systems*
*The University of Melbourne, Australia*

___________________________________________________________________

As Grids are loosely-coupled congregations of geographically distributed heterogeneous resources, the efficient utilization of the resources requires the support of a sound Performance Prediction System (PPS). The performance prediction of grid resources is helpful for both Resource Management Systems and grid users to make optimized resource usage decisions. There have been many PPS projects that span over several grid resources in several dimensions. In this paper the taxonomy for describing the PPS architecture is discussed. The taxonomy is used to categorize and identify approaches which are followed in the implementation of the existing PPSs for Grids. The taxonomy and the survey results are used to identify approaches and issues that have not been fully explored in research.

___________________________________________________________________

## 1. INTRODUCTION

As Grids are ever-changing loosely-coupled congregations of dynamic and heterogeneous resources, the efficient scheduling and allocation of resources requires the support of a sound Performance Prediction System (PPS) [1]. The performance prediction is helpful for both Resource Management System (RMS) and grid users to make optimized resource usage decisions to meet QoS requirements committed in a Service Level Agreement (SLA) [2]. For instance, if the PPS predicts a job task's future load profiles (runtimes) on the nodes of a Grid, the RMS can use such information to schedule a set of job tasks in a time optimal way. Alternatively, if the PPS can predict the cost profile of a job task on the nodes of a Grid, then an affordable set of nodes can be selected to execute a set of job tasks satisfying the user's budgetary requirements which were specified through the SLA [3].

The requirements of the PPS for a Grid span over all of the grid resources in several dimensions. They consist of the fundamentally important Application Level Prediction (ALP), namely the requirement of forecasting of the runtime of a job task on a specific machine for the given input volume, prediction of the availability of a machine and its resources for a particular duration of time, prediction of disk storage resources, prediction of network bandwidth resources, prediction of overheads of grid resources, prediction of the resultant execution time of the workflow, prediction of the reliability of grid resources, prediction of the availability of a number of nodes on a cluster/grid and so forth [2]. Therefore, the performance prediction in Grids needs to consider different approaches.

In recent times, different avenues for grid performance prediction are being explored as different research communities introduce novel approaches to perform prediction. The numerous approaches yield several different performance prediction models. Each model addresses a different performance prediction problem. Tables 1 and 2 compare a range of features of existing models, which can be used to enhance the efficiency of scheduling in grid environments.

___________________________________________________________________

Downey [4], eNANOS [5], DIMEMAS [6], Grid Performance Prediction System [7], Modelling Workloads for Grid Systems [8], QBETS [9], Smith et al. [10], Li et al. [11], Minh & Wolters [12] and GAMMA [13] focus on performance prediction of job (bag of tasks) runtime or queue time either on a cluster or parallel computers with a batch queue system, but their models can be modified to address the prediction requirements in the grid environment. The prediction approaches can be divided into two main categories. They are the prediction approaches that are based on (1) Analytical models (eg. PACE and LaPIe) and (2) Machine Learning models (eg. Smith's, Li's). An analysis of a wide range of prediction approaches is given in section 4.1.

This paper surveys through numerous PPSs that are currently available and presents taxonomy to classify them. The taxonomy covers on four different perspectives: (a) the prediction approach, (b) the resource type, (c) the resource level, (d) the job model, and they are mapped in tables 1 and 2 to selected PPSs that are designed for both clusters and grids. The main objective of this paper is to provide a basis for categorizing the existing performance prediction models for identifying future development areas such as invention of new metrics, common standards for workloads and more efficient prediction algorithms. This paper provides the reader with an understanding of the essential concepts of this research area and helps them identify outstanding issues for further investigation.

This paper consists of three major parts. The first part focuses on the general challenges for the developers of the PPSs. The second part introduces a taxonomy which separates existing PPSs. The third part includes the survey of all existing PPSs. The sections of the paper are organised as follows. Section 2 introduces the background and related work, section 3 discusses the current challenges for PPS developers and section 4 gives a description of various taxonomies for PPSs. Section 5 includes the complete survey of the existing PPSs, section 6 includes the discussion and section 7 presents a conclusion and suggestions for future work.

## 2. BACKGROUND AND RELATED WORK

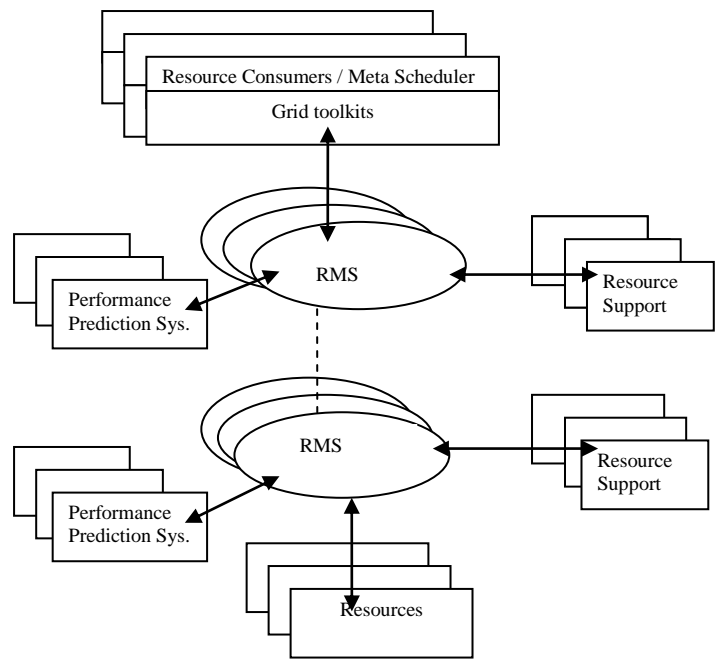

Figure 1: RMS system context

In a grid system design the RMS contains the most fundamental and essential components for its management of a Grid [3]. As a fundamental requirement, a RMS needs to have the support of sound PPS. Therefore, it is required to design or select a PPS which serves the requirement of a particular RMS [14].

In designing a large scale distributed computer system, efficient application performance and efficient system performance may require two different treatments. For instance it may not be possible for the same scheduler to optimise application performance and resource performance. One solution to this is to have two RMSs which use an application scheduler such as AppLes in conjunction with a resource scheduler such as Globus to form a two-layer RMS [15]. This suggests to us that there should be two different PPSs connected to application and resource schedulers respectively. Further, due to the diverse nature and large scale of the Grid, the resultant grid RMS is most likely an interconnection of various RMSs and each one of them needs the service of a suitable PPS. For example, the computational, data intensive and service oriented applications would require different RMS-PPS pairs and so on. Therefore, each RMS needs to have a PPS which serves its specialised requirements. Further, as the scale of the Grid grows, there can be an interconnection of various RMSs that cooperate with one another within a common framework. Figure 1 shows a block diagram of a system with multiple interconnected RMSs and their relationship to PPSs [14].

A handful of efforts has been reported for conducting surveys on Grid PPS:

Venugopal [16] has conducted a lengthy study on various taxonomies of data grids, namely *Data Grid Organization, Data Transport, Data Replication and Storage, and Resource Allocation and Scheduling*. The Data Transport, Data Replication and Storage and Resource Allocation and Scheduling Taxonomies reflect the need to have the services of sound PPSs for the prediction of network bandwidth, resources such as data storage facilities and suitable computational resources for processing data on them respectively.

The CoreGRID [17] has analysed early PPSs using a well organized template that is used to describe the prediction models and solutions. The template contains (1) Name and small description of the model, (2) Authors, (3) Scope, (4) Estimated values, (5) Predictor inputs, (6) Classes of applications or jobs, (6) Classes of resources, (7) Prediction method, (8) Prediction quality, (9) Scheduling policies, (10) Software tools, (11) Availability, (12) Architecture, (13) Support of technologies, (14) Publications, and (15) Links. Nevertheless, they consider one of the best ways to have good performance on the Grid is through performance guarantees (eg. SLAs).

They emphasise the fact that as the requirements of PPSs for the Grid span over all of the Grid resources, the prediction of resource availability needs to consider different approaches. Therefore, if a scheduler needs to have predicted levels of several different resources, that has to be done using an integrated infrastructure, and this provides access to different PPSs. Also, they suggest the PPS developers identify new performance metrics which are relevant for the Grid. They reached valuable conclusions such as the importance of the usage of data mining and AI techniques in learning prediction systems and the need for better predictors for workflow and MPI applications. They also found the necessity to have better predictors for bandwidth and data transfer rates.

Krauter [18] provides several taxonomies for RMSs with classification by *Machine Organization within the Grid, Resource Model, Dissemination Protocols, Namespace Organization, Data Store Organization, Resource Discovery, QoS Support, Scheduler Organization, Scheduler Policy, State Estimation and Scheduling Approach.* They provide a description of scheduling on the Grid in relation to their State Estimation (prediction) taxonomy which is relevant to our study.

Except for Coregrid's effort [17], in the previous surveys, Venugopal et al. [2006], and Krauter et al [2002] focus respectively on the Data Grids and RMSs and their interests in the PPSs are secondary. The CoreGrid survey provides the reader an abundance of information about early PPSs, however they do not classify the PPSs in terms of different levels of the resources, nor do they sufficiently identify the prediction approaches along the lines of analytical methodology, machine learning and *spatio-temporal correlation*, nor have they classified PPSs according to their ability to use historical information as training samples, manually or through automated means. In our survey, not only do we address a large number of PPSs, but also classify them using a number of taxonomies that is defined using above mentioned concepts, aiming to expose the missing links of the PPSs with respect to different levels of resources and applications and to motivate the researchers to invent novel prediction methods.

## 3. CHALLENGES

One of the basic challenges arises due to the heterogeneous nature presented in the grid data. This happens due to the underlying differences in the diverse type of applications, resources and different standards of grid environments. For example, the input or output data for embarrassingly distributed application is different from that of MPI application, and the data intensive applications require accessing distributed replicas which are stored across the globe.

The historical data profiles of a particular grid can be archived and would be of great help for the forecasting of future profiles. However, past experience indicates their effectiveness depends on the cleverness of their usage. Once the data is transformed into information, the next challenge is to exploit them effectively and efficiently using one of the prediction approaches. The truth is that none of the listed prediction approaches are proven 100% successful in solving the performance prediction problem. Further, it is evident that some approaches can be better used to meet a given objective than others. For example, prediction of runtimes of parallel batch job tasks in a homogeneous cluster by Modelling Workloads for Grid Systems [8]. Therefore, at this point in time, it is essential to ponder over combining several approaches to produce the best results.

The other problem is the lack of grid performance metrics. In the past, the input and output point valued parameters might have been acceptable for prediction of performance of a short job task. However, a grid job/ application consists of long job tasks and therefore such parameters may produce inaccurate results, since they can only represent a certain point of time.

Therefore, for the success of the PPSs, it is necessary to address diverse and different levels of problems which require the answers in terms of the nature of the Grid and they can be listed as follows [17].

1. Prediction of diverse parameters such as runtime, queue time, job resource requirement, resource load, communication time of MPIs, data transfer time.
2. Prediction of possible errors in the prediction of grid resources in the system. If we can collect information on possible prediction errors, then it is possible to make statistical corrections for the prediction errors of grid resources.
3. Standardization of application performance models. Though NASA has categorised the applications into different application types, still there is no common way to express application performance model.
4. There is not a standard grid workload format.
5. The PPS should be able to predict even if the input information is incomplete.
6. The PPS should be extendable for a large scope of applications and resources. In the present context and also due to the factors 3, 4 and 8, there is no single PPS to address the prediction problem of such a wide spectrum of prediction metrics and therefore it is preferable that the several prediction models may be incorporated into a single giant prediction system.
7. If possible, new performance metrics should be proposed to suit the grid. In traditional parallel computing, response time and system utilization are a major concern. In the Grid we need to reconsider traditional metrics, because the Grid is dynamic and common metrics like peak performance, throughput and point load average may be outdated or not relevant.
8. The grid environments themselves need to have single standards to be able to streamline above 3 and 4. If there is single standard, the comparison of different prediction metrics from different PPSs becomes easy. The Computational Grids, Data Grids, Service Grids etc. can be streamlined under such a single standard. Currently there are different types of grid standards, and therefore interfacing them with a certain PPS requires different strenuous adjustments to the PPS.
9. The PPS should be able to predict the quality of service.
10. The PPS should be able to predict the overheads of the system.
11. The PPS should be able to predict the availability of the required data storage.

## 4. THE TAXONOMY

Various resource types and target applications motivate the architecture of PPS. Thus PPSs have been categorized into four different taxonomies, namely prediction approach, resource type, resource level, and job model.

### 4.1. Taxonomy of Prediction Approach

The full scale simulation of activities of the grid environments has been successfully done. For example MicroGrid [19] which allows the execution of Globus applications using a virtual grid environment; SimGrid [20] which is used for simulation of "C" language application scheduling; GridSim [21] that facilitate the simulations of different classes of heterogeneous resources, users, applications, resource brokers, and schedulers in a single VO [17].

While the above effort is suitable for the representation, understanding, and analysis of the grid performance for the performance prediction of the Grid, these methods have some inherent difficulties. The main problem is that the prediction of grid commodities such as resources and services or cost needs to be done online. Predictions need to be calculated within a short period of time ($< 30s$) because in the ever-changing dynamic grid environment, the status and cost of the grid commodities are being continuously updated [17]. None of the above simulators meets these basic criteria and therefore they are not suitable for the performance prediction of the Grid.

There are two main categories of prediction methods and they are,

1. Analytical prediction models

2. Prediction models which are engaged in Machine Learning (ML) with the help of historical information.

**Analytical Prediction Models:**
In both PACE [22] and TPM [23-24] the characteristic behaviour of the job task is represented by either its code or its CPU /Disk load profile. The characteristic behaviour of the hardware environment is represented by the algorithms which mainly models the internal workings of the runnable queue and the processors. These two models belong to the school of analytical models which is developed after studying the characteristics grid application and its hardware environment. Apart from these models there are other analytical models such as LaPIe [25] which use the pLogP model to predict the overall communication time of a MPI application and the GAMMA model [13], which selects the most suitable cluster for a particular embarrassingly distributed application. The analytical models can be based on different algorithms and principles, and therefore, there can be numerous analytical models with each one having a potential for further development.

**History Based Prediction Models:**
Learning from historic information or trace data to make future predictions has always been a traditionally popular area of study. Thus, the time series modelling has been studied widely in many areas, including financial data prediction [26] and earth and ocean sciences [27].

Previous research by Wolski shows that the CPU load is strongly correlated over time, and therefore the history-based load prediction schemes are feasible [28]. This means that modelling the relationship of the historical data is of help in making accurate predictions [29-31].

Dinda conducted a complete analysis of statistical properties of host loads through a variety of load measurements collected over a wide range of time-shared machines from single PCs to clusters [29]. One of his key findings is that while load varies in complex ways, it shows high epochal behaviour. This means the pattern of change of load remains relatively constant for a relatively long period of time. The existence of epochs is significant for modelling future loads. He also found that the load does not exhibit seasonality [29]. This means that the load profile does not contain dominant underlying periodic signals on top of which are layered other signals.

Another key observation of Dinda is that the load exhibits a high degree of self-similarity with Hurst Parameter [32] ranging from 0.63 to 0.97 with a strong bias towards the top of that range. This result indicates that load varies in complex ways on all time scales and has long-range dependence and the load is difficult to model and predict [29].

Dinda carried out a thorough statistical analysis on the load traces and found that there is an opportunity to use prediction algorithms even under heavily loaded conditions. According to Dinda [33], time series analysis tools such as autocorrelation and periodogram show that the past load values have a strong influence on future values, and therefore load prediction, which is based on historical loads, is feasible and the linear time series models may be used in prediction [34].

The statistical analysis can be used on historical data to understand their behaviour. For example, Modelling workloads for grid systems [8] and Queue wait time prediction in space shared environments [4] are based on the statistical analysis of historical information. The workload modelling is introduced to make use of the collected workload traces for analysis and simulation in an analytical and manageable way [34-35]. For example, the modeller has full knowledge of the workload characteristics and therefore it is easy to know which workload parameters are correlated with each other [36]. Also, it is possible to change model parameters, one at a time, in order to investigate the influence of each one, while keeping the other parameters constant, enabling the measurement of system sensitivity against different parameters. Further, a model is not affected by policies and constraints that are particular to the site where a trace was recorded. However, the models have their own problems because its difficult to say to what degree they represent the real workloads that the system will encounter in practice [35].

Artificial intelligent techniques can be used to dynamically model historical information as it provides the basis required for the current and future behaviour of a system. For example, according to Kurowski et al. [7] mean, min, max, standard deviation, error values are calculated for each workload category. The category is decided by a template which consists of command, command argument, number of processors, maximum memory usage, host name, queue name, user name etc. Such categories with specific parameters are entered into a knowledge database as rules which are used to generate predictions for new jobs. Kurowski et al. [7] have designed their prediction system, namely the Grid Performance Prediction System (GPRES), which is based on the architecture of the expert systems.

Data mining rules can be used on historical information to find similar datasets. For example, Li uses Distance Function to categorize similar jobs and resources. The Genetic Search Algorithm is used to search for certain weights of the nearest neighbours from the historical archives. After extracting the sets of information of nearby jobs, the Instance Based Learning (IBL) prediction algorithm [37] is used for the prediction of runtimes and queue times. The details of the algorithm are found in the relevant literature [11].

The prediction models, which are based on Machine Learning, can be further divided into two major types. They are:

1. The Spatial Temporal Correlation Models (STC),

2. The models that analyse data as independent data tuples (datasets) (IDT). For example, Smith's usage of static templates to categorize data and Genetic Algorithm to search the best match. Li's use of nearest neighbours on the independent historical load profiles to categorise similar datasets (i.e. jobs with a certain similarity) [11].

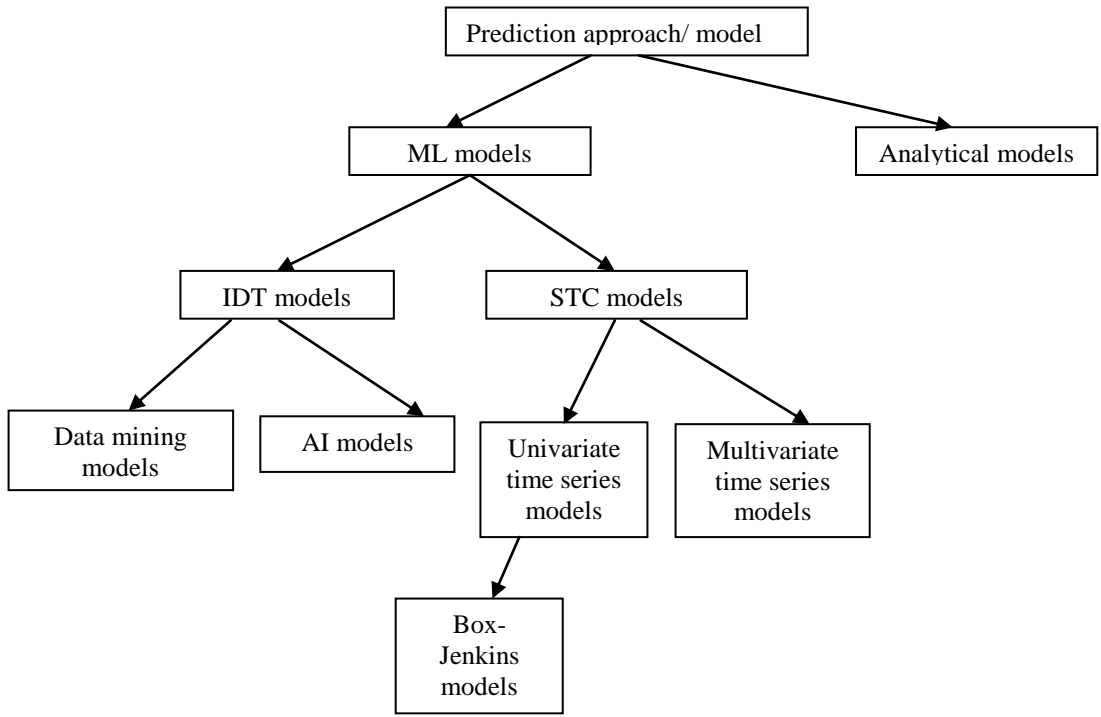

Figure 2: Prediction approach taxonomy

Figure 2 graphically presents the constituents of the Prediction approach. Another classification among the ML models is with respect to the selection of attributes/ data that would serve as training samples. The attributes/ data can be manually or automatically selected. The attributes/ data can be automatically selected, with the use of additional algorithms to select, transform, and filter.

4.2. Resource Type

There are different types of fundamental resources that are utilized by the grid users. They are (1) CPU (Processor time), (2) Memory, (3) Disk (Access cost), and (4) Network bandwidth per CPU.

The performance prediction of each fundamental resource type can be done separately as each one has different characteristics and behaviour and each resource serves a different purpose.

Other complex resource types include PC nodes, Network bandwidth and Disk storage units. A cluster can be considered as a single resource type which consists of many fundamental resource types.

The resources can be shared in two different ways, either time-shared, or space-shared. An example of a time shared resource is the manner in which the CPU in a desktop PC shares its job tasks. In this case, a few similar priority job tasks are running in round-robin fashion during their allocated time slots or time slices. In contrast, the space-shared CPU can only be allocated to a single job task at a time. The next job task may be allocated to the next available CPU. A good example of this is a Cluster computing system where a number of CPUs is managed in a space-shared manner. The Network bandwidth and Disk storage are space-shared resources [2].

As the resources can be shared in two different ways, the collected historical information differs and, therefore, the performance prediction strategies need to be different. For example, for a time-shared system, the collection of historical information is based on the load average metric because it is the future load average that needs to be predicted. In contrast, for a space-shared system it is important to predict the number of free CPUs.

The resources can be either homogeneous or heterogeneous. For example, a cluster can have identical PC-nodes which give it the homogeneous character. Also, the resource can be centralised or distributed. For example, the Grid is a loosely connected distributed heterogeneous resource. Further, the resource can be shared or dedicated. For example, a cluster of nodes, which is permanently available for HPC tasks, can be considered as a dedicated resource. On the other hand, the Grid contains a collection of PCs that are temporarily borrowed from a third party for its use and therefore considered as a shared resource. Table 2 contains the details of the PPSs which can be performed on each of the above resources.

For a particular resource, the resource consumption can be interpreted in several metrics. Therefore, it is important to measure these resources with relevant metrics that are easy to predict. For example, the CPU resource for a certain job task can be measured as (1) through CPU time share allocated to a certain job task within the elapsed second or (2) as the load average of the job task. In the OS kernel, the load average is calculated by adding the number of tasks in the CPU's queue and the number of running tasks. Therefore, it measures the load on a CPU by supplying crucial information for a prospective user who needs to decide to which nodes to submit.

### 4.3. Resource Level

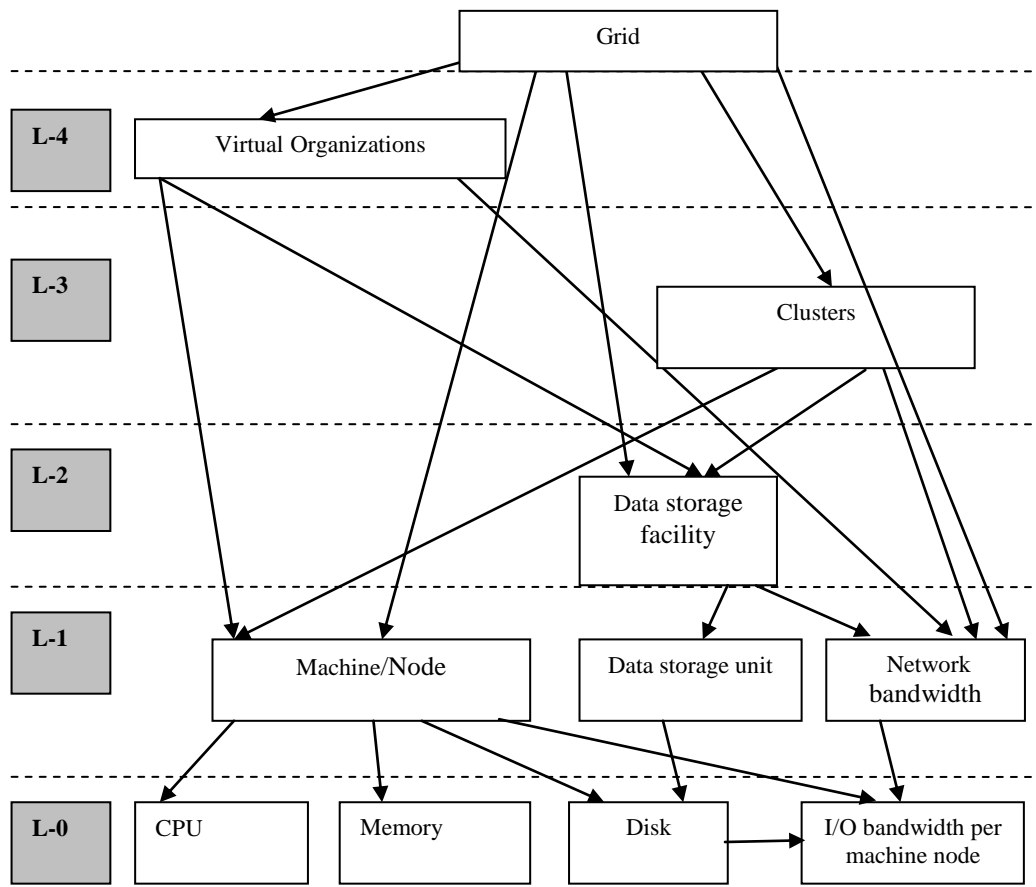

Figure 3: Resource hierarchy

Figure 3 shows the different levels of predictability on the Grid. The most fundamental resources such as CPU, Memory, Disk space and I/O bandwidth per machine node are on the ground level (L-0). Level-1 (L-1) contains Machine/ Node, Network bandwidth and a Disk storage unit. Level-2 (L-2) contains the Data storage facility. Level-3 (L-3) contains the clusters and the queue. Level-4 (L-4) contains virtual organizations and the Grid. A quite a few cluster PPSs (Table 2) which can be easily modified to predict on machine/ node resources at level-0 are considered to be level-0 predictors. Please read section 6 for more details.

**Level-0**:
**CPU:** CPU, which is available for a new job task (of a parallel application), can be predicted on a single node by using Dinda [29], Smith et al. [10], ASKALON [38], OpenSeries & StreamMiner [39], DIMEMAS [6], eNANOS [5], MWGS [8], GPRES [7], Li et al. [11], PACE [22], PPSKel [40], FREERIDE-G [41], Minh & Wolters [12], FAST [42], Prediction of Variance [43], AWP [44], or TPM [24]. This value is important because then the user knows how much further the CPU can be loaded.
**Memory:** Memory, which is available for new applications, can be predicted on a single node by using PACE [22], FAST [42] or OpenSeries & StreamMiner [39]. This value is necessary; otherwise the new application would crash without sufficient memory.
**Disk:** In FREERIDE-G project [41] the disk space access cost (time) or the data retrieval time is predicted by running the Performance Prediction Frame Work on each PC node.
**I/O Bandwidth per Node:** Both Network I/Os and Disk I/Os inherit I/O Bandwidth per node. (please see next level).
.
**Level-1:**
At this level, there are 3 main resource components that can be predicted either using the parameters of the level-0 or directly.
**Machine/ Node:** The availability of a machine/ node can be predicted directly using historically collected information as the service provider expresses the time intervals in the day that the machine is available for Grid users. NWS [28] or OpenSeries & StreamMiner [39] predicts the availability of PC nodes. Also the availability of machine/ node can be forecast after performing L-0 level predictions on CPU, memory, or disk (access cost) resources.
**Network Bandwidth:** Available bandwidth can be predicted by using NWS [28], Faerman et al. [45], PACE [22], EDG ROS [46], FREERIDE-G [41], FAST [47], Vazhkudai & Schopf [48] or PDTT [49].
**Data storage unit:** In typical data grids, a large amount of replica data is stored in different Hierarchical Storage Management (HSM) systems with access latencies ranging from seconds to hours [50]. The access latency consists of two major components and they are network access cost and storage access cost. In EDG ROS, the prediction of storage access cost is performed by CrossGrid data access estimator [46]. If the data storage consists of individual machines/ nodes, then after predicting the disk space access cost of each machine at level-0, the total data storage access cost can be calculated.

**Level-2:**
At this level, there is a single resource components that can be predicted either using the parameters of the level-0 and level-1 or directly.
**Data storage facility:** The prediction of the access cost of the Data storage facility can be done through the prediction of individual data storage units at level-1. If a data storage unit consists of several individual machines/ nodes, after predicting the disk space access cost of each machine at level-0, the total data storage access time can be calculated.

**Level-3:**
At this level there are four cluster resource components that can be predicted either using the predicted information of level-0, level-1 and level-2 or directly.
**Cluster (Parallel application's total runtime):** DIMEMAS [6] can predict the communication and computational times of a MPI parallel application. Smith et al. [10], eNANOS [5], Li et al. [11], Minh & Wolters [12] or PQR2 [51] predicts the parallel job's runtime. Also MWGS [8] or GPRES [7] predicts the parallel job's runtime.
**Cluster (Parallel application's required number of nodes):** The MWGS [8] or RBSP [52] predicts a parallel application's required number of nodes. The suitability of a parallel application to a particular cluster of nodes can be predicted using the GAMMA Model [13] therefore, it also belongs to level-3.
**Cluster (Available memory):** PQR2 or eNANOS predicts the available memory.
**Cluster (Queue wait time):** In the available PPSs, the queue waiting time is defined for a space-shared cluster of nodes and therefore it belongs to level-3. Downey [4], Smith [1999], ASKALON [38], Li et al. [11], QBETS [9] or eNANOS [5] predicts the queue waiting time. Also, MWGS or GPRES predicts the queue wait time.

**Level-4:** The suitability of a particular VO and the requirements of a particular grid need to be predicted using the predicted information of the levels below them (level-0-level-3). Sanjay & Vadhiya [53] or HIPM [54] predicts the MPI parallel job's runtime on a Grid. GIPSY predicts the parameter sweep applications runtime on a Grid. LaPIe [55] can predict the total communication time of a MPI parallel application on a Grid.

### 4.4. Job Model

Large HPC applications (or jobs) need to be grid-enabled for deployment on a Grid. New codes may be written in distributable form, but older applications that were not written for the Grid must be grid-enabled, often by being split into multiple tasks with a grid wrapper provided for each task [56]. Figure 4 shows the different levels of predictability in grid enabled job model taxonomy. At the ground level, there are grid enabled CPU bound tasks (or computational tasks), grid enabled Disk bound tasks (tasks with disk IO components) and inter task data. In the next upper level, grid enabled job tasks are related to either HPC or High Throughput Computing (HTC). The next level contains jobs and MPI parallel applications/ workflows.

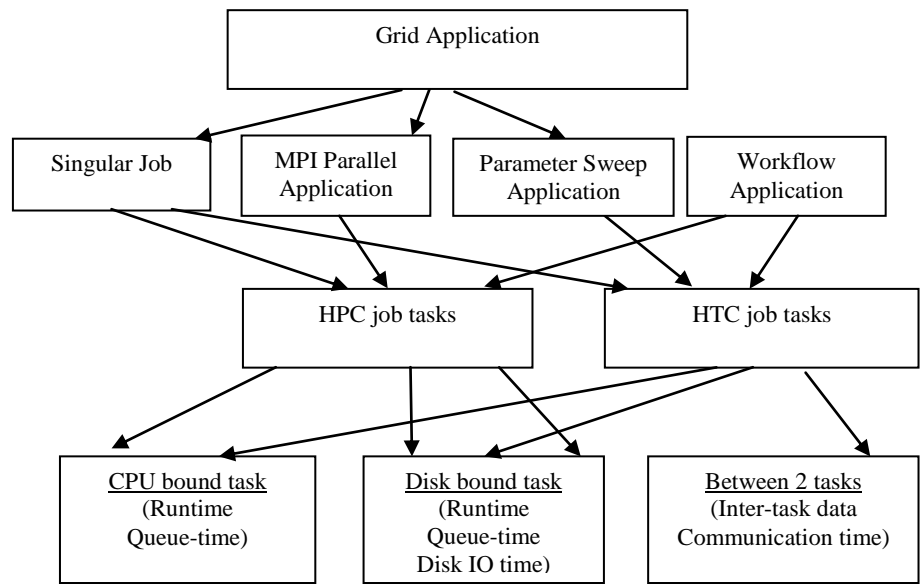

Figure 4: Grid Enabled Job model

Also, the Grid applications may consist of various groups of jobs with complex inter-task communications, therefore it is necessary to identify them through a Data Flow Graph (DFG). Such identification helps the development of generalized prediction algorithms for a particular group of MPI parallel applications or workflows. The grid enabled parallel applications are categorized into four main patterns of DFGs, in accordance with the NAS Grid Benchmarks (NGB) [57]. They are Embarrassingly Distributed (ED), Helical Chain (HC), Visualization Pipe (VP) and Mixed Bag (MB) [57]. The NGB suite is based on NAS Parallel Benchmarks (NPB) which were originally designed to provide an objective measure of the capabilities of hardware and software systems to solve computationally intensive Fluid Dynamics problems relevant to NASA [58].

At the lowest level, for the prediction of runtime of a CPU bound job task, the PPSs namely Dinda [59], Smith [10], ASKALON [38], MWGS [8], GPRES [7], eNANOS [5], OpenSeries & StreamMiner [60], PACE [1], PPSkel [40], FAST [42], AWP [44], Li [11], Minh & Wolters [12] or TPM [23, 61] can be used.

For an application that requires data transfer via the network, the PPSs, namely Faerman et al. [45] or Vazhkudai and Schopf [48], PDTT [49] or FAST [42] can be used to predict the data transfer rate. Either FREERIDE-G or EDG ROS can also be used for this purpose.

The performance prediction of a MPI parallel application/ workflow involves both computational and communication times. Therefore, it is preferable to predict the combined computational and communication time of a task rather than predicting individual components. DIMEMAS [6] or ASKALON [38] can be used for the prediction of computational and communication time of a MPI parallel application or a workflow. The MWGS [8] and GPRES [7] predict the total runtime of the parallel application.

The RBSP [52] predicts the cluster size for the MPI parallel application and the GAMMA model [13] can be used to predict the suitability of any parallel application/ workflow to a cluster.

At the highest level, HIPM [54] or Sanjay & Vadhiyar [53] can be used for the prediction of computational and communication time of a MPI parallel application on the Grid. Also LaPIe [55] predicts the total communication time of a MPI parallel application on a Grid.

## 5. SURVEY

Section 4.1 includes a summary of existing PPSs that has been proposed by researchers for various computing platforms such as clusters, grids, parallel and distributed systems, peer-peer and distributed databases. In this section, the taxonomy is used to survey some of these existing PPSs.

The PPSs for the survey are selected based on several criteria. Firstly, the survey should be concise and include a sufficient number of PPSs to demonstrate how the taxonomy can be applied effectively. Secondly, the selections of different resource types are considered for prediction. Thirdly, the selected PPSs are fairly recent work or performance prediction models that are currently in use, so that the survey creates an insight into the latest development in research.

The PPSs which are selected for the survey can be classified using several major taxonomies. As explained in section 4.1, the prediction approach can be used to classify the PPSs. The resource type and the resource level can be used for the classification of PPSs, as explained in sections 4.3 and 4.4. Table 2 provides the summary of the survey that is conducted using these taxonomies. Section 4.5 explains the job model taxonomy. The last column of Table 1 maps the job model with respect to each PPS.

Also, some of the prediction methods which have been developed for the prediction of Job runtime on a cluster can simply be modified to forecast a job task's runtime on a node of the Grid by gathering historical load profiles of similar job tasks on that particular node or similar nodes. Therefore, in the survey it is considered that these methods can be used to perform prediction of the CPU resources at level-0 of the resource tree for the Grid (Figure 3).

### 5.1. Descriptions of the Prediction Models

**Descriptions of the Analytical Models:**

**PACE** (Clusters & Grids): The PACE toolkit [22] for the performance evaluation and prediction is developed by adapting the standard methodology of the software engineering performance analysis to provide a representation of the whole system in terms of three modular components, namely the software execution module, parallelization module and the hardware module. These three modules are fed through the evaluation engine which runs the simulation of the application on a faster time scale to make required predictions.

The runtime, memory and bandwidth of the future application are the major prediction metrics that can be calculated through simulation by PACE.

Prediction quality: The average runtime prediction error is 7%.

**The Task Profiling Model (TPM)** (Clusters & Grids): The Task Profiling Model for Load Profile Prediction is proposed [23, 61], which forecasts the load profiles of job tasks of individual machines based on current and immediate past information. The Free Load Profile (FLP) or footprint of a job task on a load free node is a necessary input to the proposed Performance Prediction Model. TPM predicts the load profile (runtime) of future job tasks on the nodes of a Grid. Also, it predicts the load profiles of all currently running job tasks on each node, thereby giving an opportunity for the scheduler to protect the user set time limits of currently running job tasks. The predictions are performed by the software agents running on the nodes of a Grid. The predicted data, thus obtained, aids in choosing the most suitable set of computers for the deployment of the tasks in time-optimal manner.

Prediction quality: The average runtime prediction error is below 7%.

**DIMEMAS** (Single Machines/ Clusters)**:** The DIMEMAS [6] simulator reconstructs and predicts the total runtime which is the computation and communication time for a MPI application on a cluster. The inputs are the trace-file of the previous run of the application (a set of computation bursts and calls to the MPI primitives), description of the architecture and the model for the collective MPI primitives [17].

Prediction quality: The average prediction error versus the measured value ranges between 8% and 17%.

**LaPIe** (Clusters and Grids)**:** The LaPIe [55] predicts the overall time of a collective communication. It first subdivides the Grid network into homogenous subnets or logical clusters to handle each cluster individually and later aggregate them to form the Grid. The pLogP model is used to construct the prediction models. It consists of the communication latency –L, the message gap according to the message size –g(m) and the number of processes –P. It first establishes the performance models for a number of different communication strategies in order to select the better performing strategy for each different logical cluster. Thereafter, the best communication strategy is selected in each logical cluster. Through the analysis of the inter-clusters and intra-cluster performance predictions, it is capable of defining a communication schedule that minimises the overall execution time.

Prediction quality: The average prediction error is 5-10% for real communication experiments with varying message sizes (0- 1 MB), number of processes (1-50) and for different network infrastructures such as Fast Ethernet, Giga Ethernet and Myrinet.

**ASKALON** (Single Machines/ Clusters)**:** In the ASKALON [38], G-prophet performs the job execution and waiting time based on the minimum training set and the historical information. The execution times and different input data sizes to the job task are measured from previous job runs on a machine. If the job task with the same input data size is submitted to a similar machine, then the future execution time is predicted using the background load and memory. However, if the job task is submitted to a machine with different characteristics (eg. different CPU speed) then the new execution time on the new machine is estimated and the future execution time is predicted using the background load and memory. Mathematical models have been used in making adjustments and predictions to the execution times.

Prediction quality: The average runtime prediction error is 10%.

**GAMMA** (Clusters)**:** In the GAMMA model [13], for a particular parallel application and for a set of available clusters, a set of $\Gamma$ factors and the total functioning costs are calculated. The cluster with the least functioning cost and that best satisfies the condition $\Gamma > 1$ is selected for the deployment of the parallel application. This model is integrated into ISS VIOLA meta-scheduling environment.

A parallel application can be parameterised by the quantity,

$$\gamma_a = \frac{no.\ of\ operations\ per\ processor\ during\ the\ exec.\ of\ appli\_n\ [Mflops]}{amount\ of\ data-transfer\ by\ each\ processor\ [Mwords]} \quad (1)$$

Then a cluster can be parameterised as follows,

$$\gamma_m = \frac{peak\,performance\ of\ the\ application\ per\ processor\,[Mflops/s]}{peak\,effective\ bandwidth\,of\ the\ network\ per\ processor\,[Mwords/s]} \quad (2)$$

While the number of operations is counted in flops, the amount of data is measured in 64-bit words.

$$\Gamma = \gamma_a / \gamma_m \quad (3)$$

The cluster with the highest $\Gamma$ and the least cost may be selected for the MPI parallel application.

Prediction quality: They have experimentally demonstrated the ability to select the most suitable cluster.

**Performance Prediction with Skeletons (PPSkel)** (Clusters & Grids)**:** Sodhi et al. [40] have proposed a methodology that runs the scaled down simulation of the actual distributed system. The scaled down job tasks are automatically generated by using execution traces of CPU usage, and message exchanges. Their procedure is summarised as follows,

   a. Record job task's execution trace: The job task is executed on a controlled test bed and its execution activity, specifically CPU usage and message exchanges, are recorded.
   b. Compress execution trace into an execution signature: The repeated patterns in the recorded execution trace are used to generate a compact representation of trace by introducing loop structure. For example, if a CPU bound job task shows a flat load average curve over 4 hours, the load curve can be presented as $\alpha^4$.
   c. Generate performance skeleton program from the execution signature: The above signature is converted into a C/C++ programme that generates activity which is similar to the scaled down form of the original activity by the factor K.

The scaled down simulation is run on a particular node where the prediction of runtime needs to be performed. The evaluation of the model is done using NAS grid bench mark programs. They have successfully tested the model for both CPU bound and MPI applications.

Prediction quality: For all bench mark programmes throughout the time range 0.5s- 10s, the average error is 6.7%

**Performance Prediction Model for FREERIDE-G** (Grids)**:** Glimcher and Agrawal [41] have developed an analytical prediction model for Grid-based data mining applications. Frame work for Rapid Implementation of Data-mining Engines in the Grid (FREERIDE-G) middleware supports the high level interface for developing data mining and scientific data processing applications that involve data stored in remote repositories. Its prediction model helps to achieve the following two tasks:

   (a) Choosing the best one among multiple replicas of data
   (b) Finding the best computing resources for processing data.

This is achieved by predicting (a) data retrieval and communication times and (b) data processing times and selecting the replica and computing configuration pair where the data processing can be performed with minimum cost (time).

Prediction quality: The system performs within 5-12% when execution time is dominated by data processing time and within 4-7% when execution time is dominated by remote data retrieval time.

**Fast Agent's System Timer (FAST)** (Grids)**:** FAST [42] is a software package allowing client jobs to get an accurate forecast of communication and computation time and memory use which uses LDAP for reading and searching static data. It also uses NWS for dynamically monitoring network and hosts. The FAST dynamically acquires the CPU speed, available memory, BW, Latency, topology etc and depends on its well targeted API library which depends on linear algebraic algorithms to perform shortened calculations and combine static and dynamic data for the purpose of forecasting. The forecasts can be summarised as follows:

fast_comm_time(data_description, source, dest): The time needed to transfer the data from its location to the host on which the computation will be done.

fast_comp_time(host, problem, data_description): The time needed for the computation on a given host.

fast_comp_size(host, problem, data_description): memory space required for the problem.

fast_get_time(host, problem, data_description, localization): This function aggregates the results of others functions and forecasts the time needed to execute the given problem on a host for the data described by data_description, taking the prior localization of data and the time to get them on a host.

Prediction quality: Average prediction error is 10%

**Prediction of the QoS** (Grids)**:** Carvalho et al. [62] make an effort to predict the Quality of Service of a peer-to-peer desktop grid by determining the amount of resources available to a particular grid consumer at a certain future time $t_p$. Their model can be briefly explained as follows. In a peer-peer grid, the two peers are considered, namely donating peer $P_d$ and consumer peer $P_c$. When $P_d$ and $P_c$ interact, the resource balance with $P_d$ goes down to a minimum of zero and this guarantees that it provides a defence against whitewash attack. In contrast, the resource balance with $P_c$ goes up, proportionally to the amount of resources donated by $P_d$.

When the $P_c$ submits a bag of tasks to the Grid, it needs to gather information from each peer of the Grid. The goal is to predict the amount of resources $P_c$ is able to obtain from the Grid by the future time $t_p$. The prediction has to be performed with information gathered just prior to the submission of the tasks. Therefore, it is assumed that the balance of resource of $P_c$ and each of the $P_d$ s (all the other peers of the grid and $P_c$) do not change between the time of submission and $t_p$.

The error is defined as (ER-OR). The ER is the ratio between the estimated and requested amount of resources. The OR is the ratio between the obtained and requested amount of resources.

Prediction quality: Mean prediction error, how much of resource a peer will get from grid 7.2%

**EDG Replica Optimization Service (ROS)** (Grids)**:** The task of the ROS [46] is to select the best replica with respect to network and storage access latencies because the best replica must be accessed by an application programme. In Data Grids, a large amount of data is stored across the world in different storage systems with access latencies ranging from seconds to hours. Therefore, the data access prediction needs to be done through the cost-estimation service which consists of estimation of the access costs of the network and storage systems.

For example, if a replica is a single file,

$$File\ transfer\ \text{cost} = access\ \text{cost}\ of\ the\ network + access\ \text{cost}\ of\ the\ storage \qquad (4)$$

For estimating the access cost of the network, the EDG Network Cost-Estimation Service is used and for the access cost of the storage, the CrossGrid Data Access Estimator is used.

Prediction quality: The average estimated access cost error 6.9%

**Descriptions of ML Models: STC:**

**Wolski** (Distributed resources, clusters and grids)**:** According to Wolski, the Network Weather Service[1] (NWS) [28] provides one-step-ahead predictions for any time-series fed to its prediction module. Its prediction strategies include running average, sliding window average, last measurement, adaptive window average, adaptive window media, media filter, a-trimmed mean, stochastic gradient and autoregressive strategies. Its predicted values include CPU availability, TCP end-to-end throughput and TCP end-to-end latency.

---
[1] Wolski's NWS project provides a multifaceted prediction approach for short tasks.

Prediction quality: The CPU availability is predicted with an absolute mean error of less than 10%. The mean-based predictors are better for throughput time series. The median based predictors are better for latency time series. It is shown that mean percentage errors are less than 2.5%. The best methods have the mean percentage errors less than 0.8%.

**Dinda** (Clusters and grids)**:** On a typical shared unreserved host, Dinda [29] estimates the runtime of a computer-bound task, given the task's CPU demand and AR(16) time series prediction of the load on the host. A prediction is presented to the application/ scheduler as a confidence interval that neatly expresses the error associated with the measurement and the prediction processes-error that must be captured to make statistically valid decisions.

Prediction quality: Almost 90% of the tasks are completed in their computed confidence intervals. The target confidence interval of 95% has been used.

**Modelling Workloads for Grid Systems (MWGS)** (Parallel Computers with Batch Queue Systems)**:** Song et al. [8] use the statistical analysis and Markov-chain to predict estimation of the arrival time of a job, parallelism of a job (number of nodes) and user estimated job runtime. In their strategy, they first classify workload traces similar to Smith and then further model the workloads using Markov-chains. A Markov chain matrix has been created for each user group to model its individual behaviour and thus to predict the work loads of future jobs.

Prediction quality: The Standard Workloads are used to create the corresponding Markov chains. Thereafter using those Markov chains new workload traces are created (forecast). The KS test [63] revealed high degree of similarity. Then the model is compared with Lublin/ Feitelson model and it is found they are almost similar [64].

**Prediction of Variance** (Grids)**:** Yang et al. [65] have previously developed one-step-ahead tendency based time predictor for the prediction of CPU load as a point value. They use NWS one-step-ahead predictor for this purpose. They [43] further improved that time series based predictor to predict both mean and variance over some future time interval.

Prediction quality: 3 metrics are used to successfully compare the predictor with others. Please refer to the cited literature for details.

**Prediction of Data Transfer Time (PDTT)** (Data Grids)**:** Yang et al. [49] have predicted the data transfer times using predicted means and variances in the shared networks. They use NWS one-step-ahead predictor for this purpose. They predict the effective bandwidth using the following formulae

$$EffectiveBW = BWMean + TF * BWSD \qquad (5)$$

Where *BWMean* is the predicted mean in end-to-end bandwidth that the data will encounter during the transfer, *BWSD* is the predicted variance in end-to-end bandwidth that the data will encounter during the transfer, and *TF* is a per link *Tuning Factor* used to scale the impact of the *BWSD* on the effective bandwidth. In fact *TF* regulates the *EffectiveBW*. For example, if the variance becomes higher for a particular link, *TF* becomes lower and helps reduce the *EffectiveBW* for that link.

Prediction quality: 3 metrics are used to successfully compare PDTT with others. Please refer to the cited literature for details.

**Adaptive Workload Prediction in Confidence Window (AWP)** (Grids)**:** Wu et al. [44] have proposed a prediction methodology, Adaptive HModel (AHModel), which is based on Auto-Regression. HModel uses a fixed historical data interval as an input to predict the load n time steps ahead, within a confidence window. However, in the AHModel, the historical data interval is calculated to minimise the mean square error in the work load, before predicting the load over certain look ahead span n time steps. In other words, when the load fluctuates rapidly, the AHModel changes the historical data interval to improve the prediction accuracy.

They also use a Kalman filter to reduce the measurement errors and thereby improve the prediction accuracy. A Savitzky-Golay filter is used to smooth the spikes of the workload data in several steps of the prediction process.

Prediction quality: Mean Squared pred. error is 0.04- 0.73 as the confidence window increases from 10-50 steps.

**QBETS** (Parallel Computers with Batch Queue Systems)**:** Nurmi et al. [9] have named their method as Queue Bounds Estimation from Time Series (QBETS). They consist of four main strategies, (a) Percentile estimation, (b) Change point (of history data profile) Detection, (c) Similar job clustering (d) Machine availability interference.

They focus on space-shared batch job tasks and propose a user friendly metric to express queue wait times. According to them, the queue delay is represented to the potential users as quantified confidence bounds rather than as a specific point value prediction, because then the users can feel the probability that the job will fall outside the range. They use two metrics, namely correctness and accuracy to explain the delay. The correct prediction should be one that is greater than or equal to a job's actual queuing delay and therefore the correct predictor should be one for which the total fraction of correct predictions is greater than or equal to the success probability specified by the target percentage. The RMS error is only calculated for over-predictions as a measure of accuracy.

The other important fact is that instead of inferring from a job execution model the amount of time the job tasks will wait, a job wait time inference is made from the online actual job wait time data itself. They use time series based methods for the prediction of confidence bounds.

Prediction quality: The system predicts bounds correctly for 95% or more individual job wait times. Non-parametric Binomial percentile estimator is more effective than others.

**Descriptions of ML Models: Usage of Independent Data Tuples (IDT):**

**Downey** (Clusters)**:** Downey's statistical approach [4] predicts a job queue time in space shared environments. It is found that for cluster jobs up to 12 hours, the cumulative distribution function (CDF) of lifetimes for jobs is nearly uniform on a logarithmic x-axis [4]. Using this approximation, Downey simplifies the calculation of distribution of job lifetimes conditioned on the current age of a process. He derives formulae to calculate the median/ mean remaining lifetime of a job as a function of its current age.

For example, if there are $p$ processes running with ages $a_i$, and cluster size $n_i$ he predicts the mean/ median value of $Q(n')$ which is the time until $n'$ additional nodes become available. This calculation is straightforward because if one knows the age of a process, then using the above mentioned information he can calculate the remaining lifetime of the process [4].
Prediction quality: The average time saving per job is 13.5 minutes (Average job duration is 78 minutes). The coefficient of correlation between predicted queue times and the actual queue times from the simulated schedules is between 0.65 and 0.72.

**Smith** (Single Machine/ Cluster)**:** In previous work by Smith et al [10, 66] the historical runtimes of similar applications are used to predict the future runtime of the job in the parallel computer systems.

Smith et al [66] use a rich search technique to determine the application characteristics that yield the best definition of similarity for the purpose of making predictions. According to them, the job tasks may be judged as similar because they are submitted by the same user, at the same time, on the same computer, with the same arguments, on the same number of nodes, and so on. They use a genetic algorithm and a greedy search, looking for similar templates [67]. Eventually, they find that their technique provides more accurate runtime estimates than the techniques of other researchers. For example, they achieve mean errors that are 14 to 49 percent lower than those obtained by Gibbons [68], and 23 to 60 percent lower than those by Harchol-Balter and Downey [69].

Prediction quality: The mean prediction errors are between 40% and 59% of mean application runtimes.

Krishnaswamy [70] and Ali [71] further improve on Smith's work by introducing new techniques for data definition and search.

All of the above models assume that many different sets of templates may coexist and be applied to different environments and therefore through each template a useful job parameter can be predicted.

**Li's data mining method** (Clusters)**:** Li et al [11] present a new prediction technique for the prediction of a job task's queue wait time and runtime using the Instance Based Learning (IBL) techniques [37]. They use data mining algorithms, namely K-Nearest Neighbour and Genetic Search to find similar datasets in a history database. Li categorizes similar jobs using a distance function [72] and makes improvements on Smith et al. [10]. After extracting sets of information from nearby jobs, Instance Based Learning Algorithm is used for the prediction.

Prediction quality: The majority of jobs have relative error between -0.5 and 0.5 with the largest percentage centred around 0.

**eNANOS** (Clusters)**:** eNANOS [5] prediction service uses a set of predictors which are based on statistical techniques and data mining techniques. Similar to the predictions of Smith, eNANOS classifies the historical information of job tasks according to static templates (user, group, number of processors etc.). Therefore, the prediction of runtime and memory is done using the statistical estimators namely mean, median, linear regression and standard deviation.

The prediction of runtime can also be calculated using the clustering algorithms of the K-Nearest Neighbours, K-Means and X-Means and the above mentioned statistical estimators. In this case, instead of using a set of static templates like Smith et al. [66], for a certain job task, a set of job task typologies are created for each ß seconds using the above mentioned clustering algorithms. Thereafter, a set of predictors, based on the statistical techniques, is used on each topological group to make the predictions.

The queue waiting times are predicted using data mining algorithms of decision trees [5].

Prediction quality: In the evaluation of e-NANOS for parallel jobs, it has been compared with other RMSs and schedulers such as IBM backfilling scheduler. This system has shown a remarkable improvement in reducing runtime.

**OpenSeries & StreamMiner** (Desktop grids)**:** OpenSeries & StreamMiner framework [60] which is interfaced with the Weka 3.4 data mining library, uses data mining techniques such as Various classifiers, Decision trees, Bayesian and Support Vector Machines and AI techniques such as Fuzzy rules and Genetic algorithms for the prediction of machine loads, percentage of free virtual memory (memory load) and availability of machines.

Prediction quality: The workload and memory is predicted as a value between 0%-20%, 20%-40% and so on. In the first 2 cases Mean Squared Error is below 0.5 and the availability of PCs MSE is below 0.1.

**GPRES** (Clusters)**:** The GPRES model [7] proposes a similar approach as it groups similar jobs based on static or dynamic templates for the prediction of job's runtime, queue time and total runtime. The mean, min, max, standard deviation and error values of predicted parameters are calculated for each group. Thereafter, the groups with specific parameters are inserted into the knowledge database as rules. The reasoning system (eg. expert systems) selects rules from the Knowledge Database and generates the requested predictions.

Prediction quality: The average runtime error is 25%. The average total runtime error is 35%

**Faerman** (Collection of distributed resources)**:** Faerman et al. [45] use linear regression for prediction and combine NWS measurements with instrumentation data obtained from previous application executions to predict the data transfer performance of an application.

Prediction quality: The reported Normalized Mean Absolute Errors for file transfer throughputs using Adaptive Regression Modelling (AdRM) are less than 12%.

**Vazhkudai and Schopf** (Data Grids)**:** Vazhkudai and Schopf [48] have performed the prediction of large data transfer for efficient access of databases. The recent increase of usage of distributed databases provides an environment for the researches to share, replicate and manage access of copies of large datasets. It is important to know which replica can be accessed most efficiently. Therefore, fetching data from one of the several replica locations requires an accurate prediction of end-to-end data transfer. The large data transfers can be predicted using univariate and multivariate regression techniques. The multivariate predictors are combined with GridFTP logs, disk throughput observations and network throughput data.

Prediction quality: The univariate predictors have an error of at most 25%. When the multivariate predictors are combined with GridFTP logs, disk throughput observations and network throughput data, it provides gains up to 9% of that of univariate predictor.

**PQR2** (Clusters)**:** This machine learning prediction methodology [51] is based on PQR which is an algorithm that generates a binary tree that can combine a variety of classifiers. Matsunaga et al. [51] have further developed PQR [73] by allowing the leaves of the tree to select the best training regression algorithm from a pool of known methods. Although any regression algorithm can be placed on the pool when the fast predictions are required, the preference should be given for the parametric methods and therefore Linear-regression and Support Vector Machine (SVM) are used.

The PPS is developed to predict application's runtime, memory and disk space on a cluster using a considerable large number of applications and system specific attributes.

Prediction quality: PQR2 proved the best and required a few minutes for training /create a new model and few milliseconds to produce a single prediction. Pl. see cited material for details.

**Sanjay and Vadhiyar** (Grids)**:** Sanjay & Vadhiyar [53, 74] have developed a PPS which is suitable for tightly coupled parallel applications that are run on non-dedicated computer systems where the background load can change during the application execution. Their prediction system periodically measures the loads on the processors and network links during the executions. The total runtime which is computational and communicational is predicted using the following equation (6).

$$Total\ runtime = T(N, P, minAvgAvailCPU, minAvgAvailBW) = \frac{f_{comp}(N)}{f_{cpu}(minAvgAvailCPU)\ f_{pcomp}(P)} + \frac{f_{comm}(N)}{f_{bw}(minAvgAvailBW)\ f_{pcomm}(P)} \qquad (6)$$

Available CPUs and available BWs are measured for all processors and links at periodic intervals of time from beginning to end of the application execution. For each processor and link the average available CPU and average available BW are calculated. From these sets of values, the minimum average available CPU and minimum average available BW are calculated. The previous training samples are fed into the linear regression model to calculate the coefficients.

The various coefficients are used to represent the following effects:

The $f_{comp}(N)$ and $f_{comm}(N)$ are used respectively to represent the computational and communication complexity of the application in termes of data size.

$f_{cpu}(P)$ and $f_{bw}(P)$ are used respectively to represent the processor loads on computation and network load on communication.

$f_{pcomp}(P)$ and $f_{pcomm}(P)$ are used respectively to represent the amount of parallelism in computation and amount of parallelism in communication.

Prediction quality: In general, the average prediction error is below 30%.

**Minh & Wolters** (backfilling parallel systems)**:** Minh & Wolters [12] have improved the previous method by Li by dividing the historical jobs into big, medium, and small groups and have determined required parameters for each group to predict a job's runtime on a backfilling parallel system. The parameters include a template to categorise jobs, historical database size (*N*), the number of nearest neighbour jobs *K, the factor α*, and the factor *β*. During the training of the model the genetic algorithm is used to find the parameters.

Prediction quality: Comparing with Li, they reduced underestimated jobs by 20%, mean absolute error by 6.25% and Weighted absolute error by 17.5%.

**Hybrid Intelligent Prediction Model (HIPM)** (Grids)**:** Duan et al. [54] propose a hybrid intelligent method for performance modelling and prediction of the execution time of the workflow activity on the Grid. They combine the functionality of the neural and Bayesian networks using the methods used by Petri, to achieve high accuracy in prediction of runtimes of the workflow activities with almost negligible training times. The training time of the predictor is very low due to the introduction of the Bayesian networks. They have shown that the combined Bayesian-Radial Basis Function-Neural Network (Bayesian-RBF-NN) predictor is better than the normal RBF-NN or SVM or REP Tree.

Prediction quality: The average accuracy is 91.5%

**Regression-Based Scalability Prediction (RBSP)** (Clusters)**:** Barnes et al. [52] propose a multivariate regression to predict the ideal number of machines required for parallel application. The number of machines depends on the application, input variables, and machines under consideration. They propose three different techniques and they are as follows:
Total execution time (TOT): This technique is effective only for an application with computer-bound job tasks. The TOT calculates the total execution time and then predicts the number of machines using multivariate regression.

Maximum per-processor computation (Max): This technique is for the communication-bound applications (eg. MPI applications). The Max calculates the maximum execution time across all processors and the communication time from that same processor. This value is used to predict the number of machines using multivariate regression.

Global critical path (GCP): This technique is also for the communication-bound applications (eg. MPI applications). GCP calculates the longest execution sequence without blocking. This value is used to predict the number of machines using multivariate regression.

Prediction quality: Median prediction error is less than 13%.

**Grid Information Prediction System (GIPSY)** (Grids)**:** Verboven et al. [75] focus only on a particular class of applications, namely the parameter sweep, when developing their PPS. They select any of the following statistical/ML methods for modelling the final engine for the prediction of runtime. They are Polynomial approach [76], Radial basis functions [77], Kriging methods [78],Neural networks [79], Support vector machines [80], Nearest neighbour predictions [37] and Techniques of Iverson [81].

Prediction quality: The predictor helped the normal scheduler to improve its efficiency by 13.74%

6. ANALYSIS OF SURVEY

## 6.1. Introduction

In this section, we investigate the results of the survey. We also analyse the results of the existing prediction solutions in terms of their exploitation in grid scheduling. In the analysis of about 32 different PPSs, which are tabulated in the Tables 1 and 2, it is understood that the accurate estimation of the prediction information in the heterogeneous grid environment is a complex task.

In the survey, the majority of PPSs (17 of them) predict a job task's runtime or CPU resource. Eight PPSs predict the parallel job's queue waiting time on a cluster. Seven PPSs predict the parallel job's runtime on a cluster. A single PPS (DIMEMAS) predicts MPI parallel job's combined computational and communicational time on a cluster. Two PPSs predict the available cluster memory. Two PPSs predict MPI parallel job's combined computational and communicational time on a Grid. Eight PPSs predict only the communication time between two points. One PPS predicts the total communication time on the Grid (LaPIe). Three PPSs predict the node memory. Four PPSs predict the availability of a PC or available number of PCs. One PPS predicts the suitability of a parallel application to a cluster (GAMMA model).

Also, there is a prediction effort on Quality of Service of the resources [62]. The statuses of the current resource balance of the nodes are assumed to be the future resource balance of the nodes. Therefore this methodology can be further improved by using the historical resource balance values to predict the future resource balance of the node.

## 6.2. Meeting the Challenges

**Appropriate Performance Metrics for the Grid:**
It is necessary to introduce the metrics that are relevant to the behaviour of grid-enabled applications. In the survey, there is a handful of PPSs which introduce novel metrics. For example, instead of predicting the CPU load at a certain point in time, Yang et al. [43] have predicted the average CPU load over a certain time interval and variation of CPU load over some future time interval. They have used a set of stochastic scheduling algorithms to evaluate such predictions of future availability and variability when making resource mapping decisions.

Yang et al. [49] have proposed a Tuned Conservative Scheduling Technique that uses predicted mean and variance over the duration of the data transfer. They predict the effective bandwidth over the time interval of the data transfer using the equation (6). *TF* regulates the predicted *EffectiveBW*.

Instead of predicting a point load value ahead of n steps, Wu et al. [44] predict the load profile across the n time steps within a certain confidence window. Also, its AHModel predictor considers the fluctuations in the historical loads and changes the historical interval (w time steps) accordingly. The Mean Square Error of prediction is calculated over n and w steps, and therefore this prediction model is much relevant to the Grid.

Seneviratne & Levy [23-24, 61] use TPM to forecast the new metric individual load profiles of the future job tasks. The profiles include CPU and Disk IO intervals. The inputs to the model are the new metrics FLPs and Disk I/O maps of the currently running job tasks.

The GAMMA model [13] computes the GAMMA factor which is directly related to the efficiency of a MPI parallel application on a cluster, and helps predict the most suitable cluster for a particular MPI parallel application.

LaPIe [55] predicts using a novel metric, namely total communication time of a bag of MPI tasks or workflow on the Grid.

Also, it is relevant to include a new metric that is used by the QBETS [9] project to report the delay times on the queue. Instead of producing the average delay of recorded queue waiting times, they use the probability of past queue wait times reaching the confidence level (95%) of the predicted queue wait time to inform the clients about the prospective delays that one may have to experience. If the queue wait time exceeds its predicted value, it is considered to be an incorrect result. For example, the information that there is a 75% chance that the job task will execute within 17 minutes tells the client more about what kind of delay his job task will experience, than the information that the expected wait time of a particular job task is 30 minutes.

**Prediction of the Network Bandwidth:**
The prediction of network bandwidth has been achieved by fewer PPSs than that of job task's run time. Faerman et al. [45] have implemented the prediction of data transfer using the linear regression at resource level L-1. NWS predicts TCP end-to-end bandwidth, latency and connection time [28] at resource level L-1. NWS uses auto-regressive methods which predict the next step or data set of the collected series of data samples [28]. Therefore, for an application with long communication times, this method alone cannot provide successful predictions. Vazhkudai and Schopf [48] use multivariate regression predictors, while Yang et al. [49] predict means and variance of the network bandwidth over a certain future time interval. The latter effort is better for the Grid job tasks than the first because it predicts the result over a period of time. Since the development of NWS, there has been a good effort to predict the communication time. PACE [1] predicts the communication time of a certain job task at resource level L-1. Also, it is encouraging to see how LaPIe [55] predicts the total communication time of a bag of MPI tasks or a workflow at resource level L-4 with a reasonable accuracy. One of the major drawbacks with LaPIe is that it focuses on limited scenarios where there is no background network traffic. Also, FAST [42] predicts communication time without considering the background congestion. However, the univariate and multivariate regression predictors and methods used in NWS, take into account the history of network traffic when predicting the future BW.

**Prediction of Multiple Metrics:**
There are integrated PPSs which predict multiple parameters and they focus on predicting several diverse parameters. For example, GPRES, eNANOS, and PACE belong to this category. When the classes of applications which require prediction solutions become large, a larger number of multiple parameters may be predicted. This is applicable for grids containing heterogeneous resources where a wide range of various applications can be run. As they do not specialise in a particular resource type, their prediction accuracy may be low. However, where a grid scheduler needs to use several parameters for scheduling, it would have to use multiple Performance Prediction solutions to provide the required parameters.

**Prediction of Data Access:**
The prediction of data access time or the time required to access replicated data from any location on the Grid, consists mainly of the access time of the storage system and access time through the network. Through experience with data grids, it is known that Hierarchical Storage Management (HSM) systems are the main bottleneck rather than network links [50]. Therefore, the prediction of the access times for the HSM systems is critical to the effectiveness of the user application for reading data efficiently. In the survey, there are two PPSs for the prediction of data access. They are EDG Replica Optimization Service [46] and Prediction Model for FREERID-G [41].

6.3. Taxonomy

**Resource Type and Resource Level Taxonomies:**
There are many resource types which are required to be predicted for efficient and effective scheduling in the grid and each resource type consists of several levels. Therefore, if required, the prediction can be conducted at all the levels of a certain resource. In the resource level tree of Figure 3, the predictions which are conducted at a lower level can be transferred to a higher level without the loss of information because the nodes of the lower level are dependencies of the higher level. For example, the Level-0 predicted usage values of CPU, Memory and Disk can be used to predict the availability of a machine/node which is at Level-1. In contrast, the prediction information such as CPU capacity or Memory capacity may not be directly derived from the value predicted at Level-1. Similarly, if you predict the required number of nodes or a suitable cluster (at level-3) for a parallel application, this value cannot be transferred down the tree to extract the predicted CPU capacity, Memory capacity or Disk access cost of a particular node. The reason is that the information is tightly entangled at the upper levels of the tree and therefore accurately separating them is either extremely difficult or impossible [29, 61].

It is important that the parameter at the lowest level is fundamental and thus not a combination of any other basic parameters. Therefore, it is independent of any other metrics and it can be measured directly. Also, its characteristics and runtime environmental details can be easily understood and measured (eg. a CPU resource). Therefore, the overall result is directly related to this fundamental resource. For this reason, the accuracy of prediction of the parameter at the lower level is better than that of higher levels (eg. Dinda, PACE, PPSKel and TPM). Therefore, more predictors are developed to predict at the lowest level of the tree and transfer the results to the top level.

A resource at the higher level depends on the several types of metrics at the lowest level, and therefore its behaviour becomes complex. This makes it difficult to analyse or predict its behaviour. For example, a cluster of heterogenous nodes depends on its numerous CPUs, Memories, Disk access capabilities, and BWs which are at the lowest level of the tree (Figure 3). Therefore, it is harder to analyse or predict the behaviour of a cluster of heterogeneous nodes which is multi-dependant than a CPU or a memory.

If you compare the parameters of Table 2 and the prediction quality from each prediction method in section 4.2, you find that at the higher levels of the resource tree (Figure 3), it is difficult to make predictions accurately, and for this reason there are only a few predictors available for making predictions at the highest levels for the Grid. For example, at level-4, only Sanjay & Vadhiyar, or HIPM is capable of prediction of total runtime which is an aggregate of the computational & communication times of a MPI parallel application or a workflow.

At level-1 NWS, Faerman, EDG ROS, FREERIDE-G, FAST, Vazhkudai and Schopf, PDTT or PACE can predict the communication time between two points. NWS or OpenSeries & StreamMiner predict the availability of a machine/ node. EDG ROS or FREERIDE-G can predict the data storage access time.

At level-0, there are 17 PPSs (Dinda, OpenSeries & StreamMiner, DIMEMAS ASKALON, PACE, TPM, AWP, PPSKel, FREERIDE-G, Prediction of Variance, and FAST, *Smith*, *Li*, *Minh & Wolters*, *eNANOS*, *GPRES and MWGS*) for the prediction of CPU resource or job task's runtime. The last six PPSs are originally developed for the clusters to predict job runtime (at level-3), however they can be easily modified to predict the CPU resource at level-0. Also, available memory on a node can be predicted by either PACE or OpenSeries & StreamMiner.

Unlike a homogeneous cluster of nodes, a Grid is made of heterogeneous resources; therefore it is better to perform prediction at the lowest level because the prediction of independent resource is easy. If we try to predict a resource at a higher level, a collection of heterogeneous resources (eg. Cluster of heterogeneous nodes) are going to behave in more complex manner with several dependencies; therefore prediction is going to be complex. However, if the relevant application's historical data, its characteristics and its runtime environmental characteristics are not accessible to the predictor at the lowest level, the higher level prediction may be chosen.

**Prediction Approach Taxonomy:**
Table 1 classifies the PPSs according to the nature of their basic design, i.e., whether they belong to Analytical or ML category. The ML category has two major subdivisions and they are STC family and the group of predictors that analyses individual data sets (tuples). There are 11 analytical models, 8 ML-STC models and 13 ML-Independent Data Tuple models.

The PACE, TPM, DIMEMAS, ASKALON, PPSkel, FREEDRID-G, FAST, EDG ROS, LaPIe, GAMMA and the Prediction of QoS are the analytical models. Among the analytical PPSs, the prediction method used by PACE is based on the Software Engineering Performance Engineering methodology [22]. It uses the job tasks' source code and machine's hardware configuration to simulate the predictions. Although exposing the source code is not a popular option in the competitive IT world, this model is capable of making multiple predictions with good accuracy. For example, to predict a grid job task's runtime, memory, disk access cost and communication time with an error less than 8%, is a remarkable achievement [22]. The TPM has more in common with the prediction models that use source code and machine hardware configurations as inputs rather than history based predictors. This first category of models is not popular in the industry as exposing source code is a poor business practice. Therefore the TPM which uses FLPs instead of source code to reflect the behaviour of both CPU and disk loads could be a useful option. The other novel analytical approach is GAMMA that forecasts the level-3 resource, the suitability of a parallel application to a cluster. The novel procedure of LaPIe also attracts the attention of the reader, because it focuses only on network BW metric.

In the ML category, the automation of the collection of the training data is vital to the efficiency and effectiveness of the predictor because, for the industrial scale predictors where there are hundreds of tuples or data sets, it is not possible to collect them manually. The automation process includes extraction of data from the same prediction system or similar prediction systems and populates the real-time database with them by using awk or purl script or another software package. For parametric ML methods such as linear regression, the training data is used to calculate the model parameters and thereafter stored training data may be deleted. Usually, the process of training takes a few minutes and critics who do not want ML predictors to be part of PPSs for the Grid, point out this fact often. The positive side is that ML models are comparatively faster than analytical ones. However, for non-parametric ML methods, such as K-NN, a collection of data sets is continuously used for the prediction.

The prediction approach should be applied at the most suitable level of the resource tree (Figure 3). For example, for a cluster, job runtime and queue time can be predicted at level-3, by using Smith et al. [10, 66] or Li et al. [11] or PQR2 [51]. However we cannot apply the same strategy for a virtual organization or a Grid at the level-4 because they are a collection of heterogeneous resources (nodes). In a similar effort to predict the job runtime on a Grid, Sanjay & Vadhiyar [53] measure inputs at the level-0, therefore this method belongs to the level-0 as much as to the level-4. Therefore for a Grid we need to apply these methods at level-0 to separately predict job task's runtime on each node.

The only exception to this is that when HIPM [54] makes an effort to predict the job runtime on the Grid at the level-4. It measures the inputs at the level-4, for example its inputs are Activity type, Activity name, Arguments, Queue time, Execution time, Number of processors etc. The problem is that the Grid is dynamic and heterogenous collection of resources (nodes), therefore considering it as a single entity may be a mistake because unlike a cluster, it is a complex system.

### 6.4. Prediction Models and Grid Scheduling

In this section, the existing performance prediction solutions are investigated in terms of their application in grid scheduling. To this end, firstly the models and their usability for grid scheduling are studied. Secondly the availability of software is investigated.

**Applicability**:
The requirements of the PPSs result from the heterogeneity and dynamic nature of the grids. The main requirements that must be taken into consideration during the analysis of models applicability are listed below.

1. Prediction of important parameters. (such as runtime, queue waiting time job resource requirement, resource load, communication time, data transfer time.)
2. Information about errors.
3. Performance guarantees and /or small prediction errors.
4. Performance prediction based on incomplete and/or imprecise information.
5. Large scope of handled applications and resources.

The parameters predicted by PPSs which are included in the survey are presented in the 4[th] column, under predicted metrics of Table 2. The information, the number of PPS solutions that support the particular parameters is summarised in section 6.1.

If a scheduler needs to use multiple parameters for scheduling, it would have to employee multiple performance prediction solutions to obtain required predicted parameters. In section 6.2 under Prediction of Multiple Metrics, this solution has been discussed. Similar conclusions have been made based on the analysis of different classes of applications and their input parameters handled by the considered solutions.

There are prediction solutions which are more generic and provide predictions of great number of parameters for a wide class of applications. For an example, GPRES, eNANOS [5, 7]. However generic solutions that take general input parameters from available data archives can be relatively poor when compared with specialised performance prediction models which require more detailed description of application and resources. Those latter prediction models include DIMEMAS and Gamma [82-83].

**Availability of Software:**

In addition to requirements concerning performance prediction models themselves, there are other practical issues such as that influence the usability of particular solution. They are

1. Software availability
2. Availability of easy to use and generic interface.
3. Ability to integrate with common resource management and grid technologies.

There are a few proficient software packages available such as NWS [28], the Intelligent Grid Scheduling System (ISS) which implements the Gamma model [83] and Grid-Prophet which implements ASKOLON [38]. But most of them are not sufficiently mature to satisfy the customer's/user's specific needs. This is a major hurdle in taking advantage from available PPS. Therefore the above mentioned problems need to be discussed with the software creators because the licensing rights for versioning and maintaining of each PPS software belong to them.

| Performance prediction model | Category of the predict model | | | | | Input metrics/training data | Res. type | Class of job (time / space shared) |
|---|---|---|---|---|---|---|---|---|
| | Machine learning model | | | | Analytical model | | | |
| | STC | IDT | Selection of data (Man. / Auto.) | | | | | |
| | | | Man. | Auto. | | | | |
| Downey | | Yes | | Yes | | Jobs' historical runtimes, queue times, cluster size and details of their processors are collected from the similar sites to plot the distribution of total allocation time of jobs in log space. | Queue | Parallel jobs (space) |
| Dinda | Yes | | | Yes | | (1) Historical host load data is automatically stored. (2) Free load runtime of the new job task needs to be recorded. | CPU | short j. tasks (100ms-10s) (time) |
| NWS | Yes | | | Yes | | Historical values of CPU usage, Memory usage, TCP end-to-end bandwidth and latency, and connection time are stored automatically. | CPU availability, BW | parallel j tasks (time) |
| Faerman | Yes | | | Yes | | Historical information of data transfer. NWS measurements such as TCP end-to-end bandwidth and latency and the connection time are automatically stored. | BW | Appli. with data transfer infor. (time) |
| Vazhkudai & Schopf | Yes | | | Yes | | GridFTP logs, disk throughput observations and network throughput data are automatically stored. | BW | Appli. with data transfer infor. (time) |
| Smith | | Yes | | Yes | | Sets of template attributes and their relevant profiles of historical workload are automatically stored. | CPU, Queue | Parallel job (space) |
| DIMEMAS | | | | | Yes | Sets of computation bursts & calls to MPI primitives. Description of the application architecture. | CPU & BW | MPI parallel jobs (space) |
| Prediction of Variance | Yes | | | Yes | | CPU load time series is automatically and online stored on each node. | CPU | Parallel jobs (time) |
| LaPIe | | | | | Yes | Communication latency, message gap according to message size and number of processors. | BW | MPI parallel jobs (time) |
| ASKALON | | Yes | | Yes | | Job tasks' historical information such as, job task names, runtimes, input parameter sizes, processor speeds are automatically stored from previous actual runs. If the historical information is not available then it is manually supplied from one of the identical machines. | CPU, CPU (total) & Queue | Parallel jobs (time) |

Table 1: Classification of Performance Prediction Approaches…………

| Performance prediction model | Category of the predict model | | | | Input metrics/training data | Res. type | Class of job (time / space shared) |
|---|---|---|---|---|---|---|---|
| | Machine learning model | | | Analytical model | | | |
| | STC | IDT | Selection of data (Man./ Auto.) | | | | |
| | | | Man. | Auto. | | | |
| Li | | Yes | | Yes | The profiles of historical workloads are automatically stored. | CPU, Queue | Parallel jobs (space) |
| PDTT | Yes | | | Yes | Network bandwidth time series is automatically and online recorded at constant width time intervals. | BW | Data intensive appli. (time) |
| eNANOS | | Yes | | Yes | Both statistical and data mining predictors need the same inputs: job name, user name, group name, no. of processors, job name & script name, they are automatically stored as load profiles. | CPU, Memory, Queue | MPI parallel jobs (space) |
| OpenSeries & StreamMiner | | Yes | | Yes | Historical values of workload such as CPU idleness, percentage of free virtual memory, machine availability (switched on or off) and user presence indicator (logged on or off). The attribute selection process which has 3 phases can be semi-automated. | CPU, Memory & CPU availability | Parallel jobs (time) |
| GPRES | | Yes | | Yes | The historic jobs are categorised according to static or dynamic template attributes. Then the mean of values estimated parameters are calculated for each category. Such categories with a specific set of values are inserted into the knowledge database as rules. This process can be automated | CPU, Queue | Parallel jobs (space) |
| MWGS | Yes | | | Yes | User-name, submission time, job ID, number of nodes requested, user-predicted job runtime, actual job runtime are the input to generate two Markov chains for the runtimes & num of nodes. The data collection and the generation process can be automated. | Queue, No. of nodes, CPU (total) | Parallel jobs (space) |
| GAMMA | | | | | Yes | Computational and network traffic information of the application and the cluster and costing parameters | Cluster | Parallel job (space) |
| PACE | | | | | Yes | Software code of the applications and machine and environmental details. | CPU, Mem, BW, Queue | MPI parallel job (space/ time) |

Table 1: Classification of Performance Prediction Approaches…………

| Performance prediction model | Category of the predict model | | | | Input metrics/training data | Res. type | Class of job (time / space shared) |
|---|---|---|---|---|---|---|---|
| | Machine learning model | | | Analytical model | | | |
| | STC | IDT | Selection of data (Manual/ Automated) | | | | |
| | | | Man. | Auto. | | | |
| TPM | | | | | Yes | FLPs of the job tasks and machine environmental details. Disk IO maps of the job tasks. | CPU, Disk | Parallel job (space/ time) |
| PPSkel | | | | | Yes | The records of execution activities of the CPU usage, Memory consumption and MPI message exchanges are taken from the same program. | CPU, BW Mem., | MPI parallel jobs (time) |
| EDG ROS | | | | | Yes | Characteristic details of CPU, network and storage. | Disk, BW | Remote data processing appli. (time) |
| Sanjay & Vadhiyar | | Yes | | Yes | | Available CPUs & Available BWs are automatically measured for all processors and links at periodic intervals. The calculated coefficients are used in the total runtime equation. | CPU & BW | MPI parallel jobs (time) |
| PQR2 | | Yes | | Yes | | Application and system-specific attributes such as cluster name, CPU clock, amount of memory, location of data, CPU speed, memory speed, disk speed, number of threads. The data sets can be collected automatically. | CPU, Mem, Disk | MPI parallel jobs (space) |
| QBETS | Yes | | | | | Historical data profiles of similar jobs. The data is collected online and the collection process can be automated. | Queue | Parallel jobs (space) |
| FREERIDE-G | | | | | Yes | No. of storage nodes, dataset size, network bandwidth, execution speed, disk speed, no. of computing nodes, & the corresponding values of the outputs. | Disk, BW, CPU | Remote data processing appli. (space) |
| Minh & Wolters | | Yes | | Yes | | Original inputs: user_name, group_name, queue_name, job_name, Intermediate parameters: historical database size (N), no of nearest neighbour jobs K, the factor α and β. Training parameters: user_name, group_name, queue_name, job_name, point_of _separate, N, K, α, β. The traces are collected automatically and training parameters are calculated automatically. | CPU | Parallel jobs (space) |

Table 1: The Classification of Performance Prediction Approaches…………………..

| Performance prediction model | Category of the predict model | | | | Input metrics/training data | Res. type | Class of job (time / space shared) |
|---|---|---|---|---|---|---|---|
| | Machine learning model | | | Analytical model | | | |
| | STC | IDT | Selection of data (Manual/ Automated) | | | | |
| | | | Man. | Auto. | | | |
| HIPM | | Yes | | Yes | Data/ Activities of the workflow application, such as type (eg. metric multiplication), name, arguments, problem size, preparation time, user name, grid site, submission time, queue time, external load, processors, execution time can be automatically collected and the predictor can be trained fast using Bayesian network. | CPU & BW | Workflow job (time) |
| RBSP | | Yes | | Yes | Job's input variables, number of machines under consideration are inputs to the regression equation. The collection of the training data can be automated. | Cluster | MPI Parallel jobs (time) |
| AWP | Yes | | | Yes | Historical workload points are collected online | CPU | Parallel jobs (time) |
| FAST | | | | | Yes | Dynamically collected data such as CPU speed, workload, BW, available memory, batch system. | CPU, Memory, BW | Parallel jobs (space) |
| GIPSY | | Yes | | Yes | (a) Initial Training sample selection can be automated. (b) Selection of the model may be automated subjected to the condition. The selected models can be run until the predicted runtime error converges, and then the most suitable model will be selected.. | CPU | Parameter sweep jobs (space) |
| Prediction of the QoS | | | | | Yes | The current measurement of the balance of resources of all the nodes (CPU, Mem etc). | CPU, memory | Parallel jobs (space ) |

Table 1: The Classification of Performance Prediction Approaches

| Prediction model | Res. type | Res. level | Predicted metrics | Centralized/ Decentralised | Homo- geneous/ Hetero- geneous | Dedicated/ Shared |
|---|---|---|---|---|---|---|
| Downey | Queue | L-3 | Queue time | Centralized | Homo | Dedicated |
| Dinda | CPU | L-0<br>L-0 | Host load,<br>Job task's runtime | both | both | both |
| NWS | CPU availability, BW, | L-1<br>L-1<br>L-1 | CPU availability,<br>TCP end-to-end throughput,<br>TCP end-to-end latency. | Decentralised | Hetero | Shared |
| Faerman | BW | L-1 | Data transfer rate. | Decentralised | Hetero | Shared |
| Vazhkudai and Schopf | BW | L-1 | Data transfer rate. | Decentralised | Hetero | Shared |
| Smith | CPU<br>Queue | L-3<br>L-3 | Job's runtime,<br>Queue time | Centralized | Homo | Dedicated |
| DIMEMAS | CPU & BW | L-3 | Job's runtime, | Centralized | Homo | Dedicated |
| Prediction of Variance | CPU | L-0 | CPU load mean & variance over a time | Decentralised | Hetero | Shared |
| LaPIe | BW | L-4 | MPI job's communication makes span | Decentralised | Hetero | Shared |
| ASKALON | CPU,<br>CPU (total) &<br>Queue | L-0<br>L-3<br>L-3 | Job task's runtime,<br>Job's runtime,<br>Queue time. | Centralized | Hetero | Dedicated |
| Li | CPU,<br>Queue | L-3<br>L-3 | Job's runtime,<br>Queue time. | Centralized | Homo | Dedicated |
| PDTT | BW | L-1 | Data transfer time between 2 nodes | Decentralised | Hetero | Shared |
| eNANOS | CPU,<br>Memory,<br>Queue | L-3,<br>L-3,<br>L-3 | Job's runtime,<br>Memory,<br>Queue time | Centralized | Homo | Dedicated |
| OpenSeries & StreamMiner | CPU,<br>Memory,<br>CPU availability | L-0<br>L-0<br>L-1 | Idle % of CPU,<br>Memory,<br>Availability of PCs | Decentralised | Hetero | Shared |
| GPRES | CPU,<br>Queue | L-3<br>L-3 | Job's runtime,<br>Queue time | Centralized | Homo | Dedicated |

Table 2: The classification of the Resource Types…………

| Prediction model | Res. type | Res. level | Predicted metrics | Centralized/ decentralised | Homo-geneous/ Hetero-geneous | Dedicated/ Shared |
|---|---|---|---|---|---|---|
| MWGS | Queue, No. of nodes, CPU (total) | L-3 L-3 L-3 | The arrival time of job, No. of nodes, Job's runtime | Centralized | Homo- | Dedicated |
| GAMMA Model | Cluster | L-3 | 1. For each cluster $\Gamma$ ($\gamma_a / \gamma_m$) 2. Total usage cost. | Centralized | Homo- | Dedicated |
| PACE | CPU, Memory, BW, Queue | L-0, L-0, L-1, L-3 | Job task's runtime, Memory, Communication. Time, Queue time. | Decentralised | Hetero | Shared |
| TPM | CPU, Disk | L-0 L-0 | Load profiles of future Job tasks. Disk access time | Decentralised | Hetero | Shared |
| PPSkel | CPU, BW, Memory, | L-0, L-0, L-0 | MPI Job task's runtime (CPU, communication and memory). | Decentralised | Hetero | Shared |
| EDG ROS | Disk, BW | L-1 L-1 | Data retrieval time & communi. Time | Decentralised | Hetero | Shared |
| Sanjay & Vadhiyar | CPU & BW | L-4 | Job's runtime | Decentralised | Hetero | Shared |
| PQR2 | CPU, Memory, Disk | L-3, L-3, L-3, | Job's runtime, Memory, Disk space | Decentralised | Hetero | Shared |
| QBETS | Queue | L-3 | Probability of past queue wait times reaching the confidence level (95%) of the predicted queue wait times and the RMS error of job tasks that delays less than the predicted value | Centralized | Homo | Dedicated |
| Prediction Model for FREERIDE-G | Disk, BW, CPU | L-0, L-1, L-0, | Data retrieval time, commun. time, & data processing time. | Decentralised | Hetero | Shared |
| Minh & Wolters | CPU | L-3 | Job's runtime | Centralized | Homo | Dedicated |
| HIPM | CPU & BW | L-4 | Job's runtime | Decentralised | Hetero | Shared |
| RBSP | Cluster | L-3 | No of machines | Centralized | Homo | Dedicated |
| AWP | CPU | L-0 | Load profile | Decentralised | Hetero | Shared |
| FAST | CPU, Memory, BW | L-0, L-0 L-1 | Processing runtime, Memory, Communication time | Decentralised | Hetero | Shared |
| GIPSY | CPU | L-4 | Job's runtime | Decentralised | Hetero | Shared |
| Prediction. of QoS | CPU, Mem etc | L-0 | Future available balance of resources (CPU, Mem etc) | Decentralised | Hetero | Shared |

Table 2: The classification of the Resource Types

## 7. CONCLUSIONS

In this paper, taxonomy is proposed to characterize and categorize various aspects of PPSs that can support the preparation of efficient and effective application schedules for the Grid. The taxonomy covers four different perspectives: (a) the prediction approach, (b) the resource type, (c) the resource level and (d) the grid enabled job model. A survey is also conducted where the taxonomy is mapped in Tables 1 and 2 to selected PPSs that are designed for both clusters and grids. The prediction approach taxonomy is used to identify certain characteristics of machine learning and analytical models and their critical parameters are mapped in Table 1. The application model taxonomy separates various levels of a grid enabled application and it is mapped in the last column of Table 1. The resource type and resource level taxonomies have been used to identify the PPSs that are implemented at various levels of the same resource (Table 2). The survey is helpful to us to analyse the gap between what is already available in existing PPSs and what is still required so that we can identify the research requirements which can be implemented in future projects.

When one compares the identified expectations in section 3 with the survey results of the PPSs, there are only few achievements of which one can be proud. Dinda, PACE, TPM, DIMEMAS, ASKALON, OpenSeries & StreamMiner, PPSkel, AWP or FAST make the predictions of CPU resource at the level-0 and can be considered as the most reliable.

Prediction of a parallel job's runtime and queue waiting time (level-3) on a cluster can be achieved by several predictors. It is not possible to use the same methodology on a Virtual Organization or a Grid (level-4) because of their heterogeneity. However, it is explained that the predictors which predict the job's runtime on a cluster can be modified to predict a job task's runtime on a node (level-0) and therefore, there are 17 predictors (level-0) to forecast CPU resource on a node of the Grid. Also it is evident from the survey & taxonomies that the prediction on a Grid, should preferably be done at the lowest level of the tree (Figure 3). Thereafter the prediction results of each node can be transferred from the lowest level to the highest level of the resource tree (Figure 3).

The usage of regression based techniques for the prediction of data transfer times has become a fair trend. Yang et al. [49] is based on the NWS [28] predictor, making predictions over a certain time duration, making it more relevant than others [48] for the Grid. Steffenel [55] has taken a different approach for predicting the most suitable BW for a MPI parallel application on the Grid. His methodology namely LaPIe [55] helps to select the scheduling strategy to minimise the overall communication time on the Grid.

Although a few new performance metrics are proposed by some PPSs, the same old metrics are used by many PPSs because their underlying algorithms are not changed to capture some of the vital characteristics of Grid applications. In contrast, in the TPM, the metric FLP captures total behaviour of the job tasks. Yang et al. [43, 49], AWP, GAMMA and LaPIe can be identified as other initial contributors in this area.

The other major problem is the lack of global standards for the Grid. Under such common standards, both application performance model and workload formats can be categorised. Currently, NASA grid bench marks have introduced some application types and this may be a good starting point [57].

The other important need is to have a single framework which addresses the prediction of several different parameters. Currently, there are few PPSs which have a limited capability of doing so (eg. GPRES, eNANOS, and PACE). However, the major problem that would arise from such a massive integration is that the bigger the framework the more complex and slower it will be, because at this point of time such a single framework has to be developed by integrating several different PPSs.

Also the prediction of required data storage for a certain application requires the assistance of PPSs.

None of the PPSs has made an effort to forecast prediction errors, nor have they attempted to perform the prediction when their required information is incomplete. There are no reliable PPSs for the prediction of overheads of the Grid. Therefore, these three areas still remain wide open for the researchers who explore further and evaluate how influential their results would be for the completion of good grid schedules.

**Acknowledgement**
We thank Srikumar Venugopal, Anton Beloglazov, Nikolay Grozev and Adel Nadjaran Toosi for their valuable comments which have enriched the content of this paper.